\documentclass[
 reprint,
 aps,
 prapplied,
 amsmath,amssymb,
 floatfix,superscriptaddress
]{revtex4-2}

\usepackage{graphicx}
\usepackage{dcolumn}
\usepackage{bm}

\usepackage{ulem}
\usepackage{bm}
\usepackage{xcolor}
\usepackage{hyperref}

\begin{document}

\preprint{APS/123-QED}

\title{Towards Schrödinger Cat States in the Second Harmonic Generation}

\author{Ranjit Singh}
\email{ranjit.singh@mail.ru}
\affiliation{Independent Researcher, Domodedovo, 142000, Moscow region, Russia}

\author{Leonid A. Barinov}
\affiliation{Moscow Institute of Physics and Technology,
9 Institutsky Pereulok, Dolgoprudny, 141701, Moscow Region, Russia}
\affiliation{Russian Quantum Center,
Skolkovo Innovation Center, Bolshoi Boulevard 30, 121205, Moscow, Russia}

\author{Grigori G. Amosov}
\affiliation{Department of Probability Theory and Mathematical Statistics,
Steklov Mathematical Institute of Russian Academy of Sciences,
8 Gubkina St., Moscow, 119991, Russia}

\author{Anatoly V. Masalov}
\affiliation{Moscow Institute of Physics and Technology,
9 Institutsky Pereulok, Dolgoprudny, 141701, Moscow Region, Russia}
\affiliation{Russian Quantum Center,
Skolkovo Innovation Center, Bolshoi Boulevard 30, 121205, Moscow, Russia}
\affiliation{P.N. Lebedev Physical Institute of the Russian Academy of Sciences,
Leninsky prosp. 53, 119991 Moscow, Russia}

\date{\today}

\begin{abstract}
We investigate the quantum evolution of the pump field in second-harmonic generation under strong pump depletion. Starting from a coherent state, the pump develops a nonclassical phase-space structure resembling a Schrödinger cat state. This behavior originates from phase instability induced by vacuum fluctuations of the harmonic mode. 
A rigorous quantum analysis has been performed for mean photon numbers up to $\langle \hat n \rangle = 100$ in pump mode. For larger photon numbers, up to $\langle \hat n \rangle = 10^{7}$, the dynamics have been analyzed using a classical trajectory method with sampled initial conditions that reproduces the main features of the quantum evolution. The results indicate that nonlinear frequency conversion can generate macroscopic superposition-like states of the pump field. Although the resulting state is not pure due to correlations with the second-harmonic wave, it remains non-classical with negative zones of Wigner function. 
These results indicate that strongly nonlinear frequency conversion can provide a scalable route toward macroscopic nonclassical states of light.
\end{abstract}

\maketitle


\section{Introduction}
	Light in a Schrödinger cat (SC) state is in demand for quantum technology applications \cite{Vlastakis2013,Grimm2020}. It can be used to implement quantum encoding protocols with continuous variables \cite{Jeong2002,Ralph2003,Lund2004,Mirrahimi2014,Joshi2021}. These are states of the type \(c_1|\alpha\rangle + c_2|-\alpha\rangle\) with coefficients of similar magnitude \(|c_1| \approx |c_2|\). A process leading to the formation of a Schrödinger cat state from a coherent light is known. It is based on nonlinear optical conversion of light in a medium with Kerr nonlinearity \cite{Miranowicz1990}. Implementation of this process is difficult due to increased requirements for radiation intensity in cubic nonlinear media. As estimates show (Appendix 1), to achieve such generation, the nonlinear phase accumulated in the Kerr medium must reach the enormously large value \(\Phi_{\text{NL}} = \pi\langle n\rangle\), where \(\Phi_{\text{NL}} = \frac{\omega}{c}n_2 Iz\), and \(\langle n\rangle\) is the mean photon number of the input beam,  $n_2$ - nonlinear refraction coefficient, $z$ - length of medium. We note that the "nonlinear phase" $\Phi_{\text{NL}}$, which is well-known in classical descriptions, has the meaning of evolution time in quantum calculations. Only initial stage of Kerr nonlinearity evolution can be used to generate sub-Poissonian light \cite{Singh2025}.

A different approach to generating Schrödinger cat states has been proposed in papers \cite{Singh2024arXiv, Singh2026, Gorshenin2025}, where the target state is formed in the parametric wave during the process of parametric down-conversion.

In modern practice, Schrödinger cat states have only been realized with small amplitudes (\(|\alpha| < 2\)) using conversion of few-photon light states \cite{Ourjoumtsev2006,Sychev2017}. 
    

    In the present paper we consider an alternative possibility for forming Schrödinger cat states during second harmonic generation, where the target state is formed at the pump frequency. The resulting state has an amplitude comparable to the initial one. This method uses the bifurcation instability of the pump wave phase during second harmonic generation (Appendix 2).
	
	The formation of the quantum state of the pump mode during second harmonic generation occurs under conditions of phase instability. While in the classical picture of the pump field evolution there is no phase instability, the pump wave evolution in quantum regime suffer instability due to nonzero vacuum field of second harmonic input: some components of vacuum field induces "plus" phase shift of pump field, and opposite components induces "minus" shift.
     As a result, the pump field state acquires features of a superposition of fields with opposite phases, i.e. becomes similar to a Schrödinger cat state. The possibility of this effect was indicated in \cite{Nikitin1991}.
	
\section{Quantum Model}
	
	We consider the evolution of the quantum state of the pump wave during second optical harmonic generation according to the Schrödinger equation with the Hamiltonian
	\begin{eqnarray}
		\hat{H} = -g \hbar \left( \hat{a}_1^{\dagger 2} \hat{a}_2 + \hat{a}_1^{2} \hat{a}_2^\dagger  \right),\label{Hint}
	\end{eqnarray}
		where $g$ is the nonlinear coupling constant for collinearly propagating interacting waves, proportional to the second-order susceptibility $\chi^{(2)}$ and operators $\hat{a}_1$ and $\hat{a}_2$ are the pump wave and second harmonic operators, respectively. Here it is assumed that the wave interaction occurs under perfect phase matching conditions. The evolution operator has the form
		\begin{eqnarray}
	\hat{U} = \exp \left\{ i g t\left( \hat{a}_1^{\dagger 2} \hat{a}_2 + \hat{a}_1^2 \hat{a}_2^{\dagger} \right) \right\}.\label{Uop}
	\end{eqnarray}
	The evolution (\ref{Uop}) converts the two-mode state matrix elements
	\begin{eqnarray}
	\hat{a}_1^{\dagger 2} \hat{a}_2 |n_1, n_2\rangle = \sqrt{(n_1 + 1)(n_1 + 2)n_2} |n_1 + 2, n_2 - 1\rangle,\\
	\hat{a}_1^2 \hat{a}_2^{\dagger} |n_1, n_2\rangle = \sqrt{n_1(n_1 - 1)(n_2 + 1)} |n_1 - 2, n_2 + 1\rangle.\label{HintMatEle}
	\end{eqnarray}
		In the calculations, the pump field is initially considered to be in a coherent state, and the second harmonic in a vacuum state. The Schrödinger equation for this problem has been solved within the limits of pump photon number $\langle n\rangle\leq $ 100. 
        The behavior of the mean photon number of the pump wave versus interaction time is shown in Fig.~\ref{fig:fig1}. The mean photon numbers calculated by the classical simulation method (see below) are also shown.

\begin{figure*}[htbp]
    \centering
    \includegraphics[width=\linewidth]{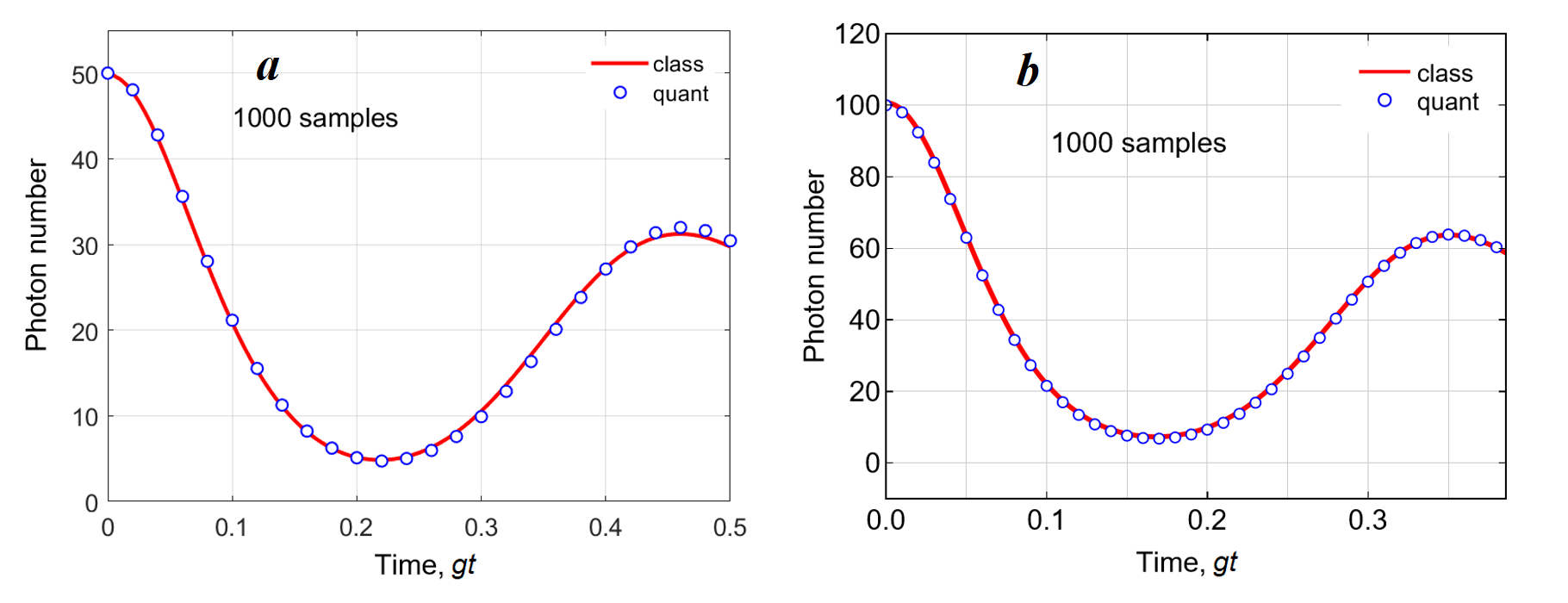}
    \caption{Dependence of the mean pump wave photon number on dimensionless interaction time $gt$: circles — quantum  calculations, red line — calculations by classical simulation method; {\it{a}} — initial photon number 50, {\it{b}} — initial photon number 100. The classical simulation used averaging over 1000 trajectories (samples).}
    \label{fig:fig1}
\end{figure*}
	
\begin{figure*}[htbp]
		\centering
	\includegraphics[width=\linewidth]{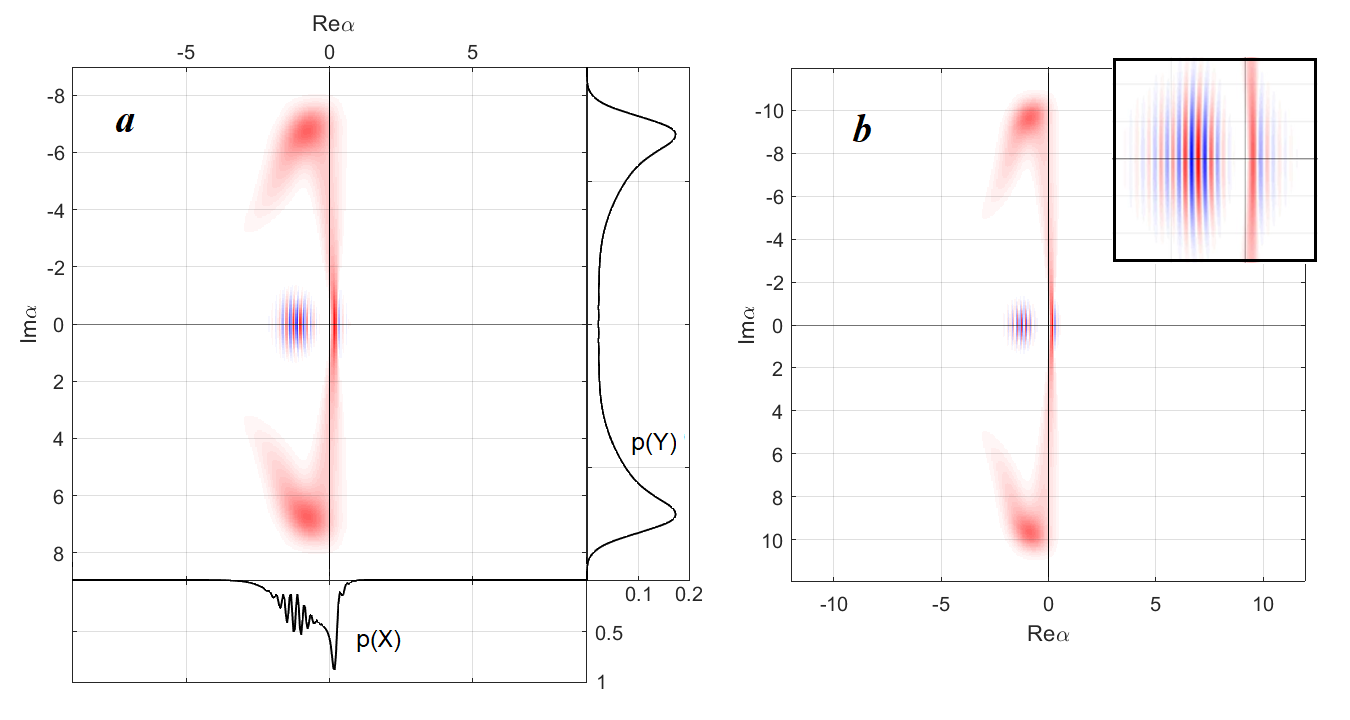}
		\caption{Wigner quasiprobability for the pump wave: {\it{a}} — initial photon number 50, time $gt =$  0.463 (Fig. 2{\it{a}}), quadrature probability densities are presented on the sides; {\it{b}} — initial photon number 100, time $gt =$ 0.352 (Fig. 2{\it{b}}), inset shows the central part of the Wigner function.}
		\label{fig:fig2}
	\end{figure*}

	The mean photon number exhibits non-monotonic behavior, despite perfect phase matching; this is a feature of quantum behavior. Parallel analysis of the Wigner quasiprobability function for the pump wave has shown that the pump state during evolution first demonstrates squeezing, and then approaches a Schrödinger cat state: the best squeezing corresponds to the first minimum of photon number, while the SC state is formed at the first maximum after depletion. The corresponding quasidistributions are shown in Fig.~\ref{fig:fig2} at the times of the first maximum after depletion. Two distinct zones of quasiprobability are formed, and an interference structure is also present. The presence of interference indicates coherence of the distinct field states. Despite some difference between the calculated states and ideal SC states, the main features of the SC state are reproduced. The resulting state of the pump mode is mixed because of entanglement with the second-harmonic mode. In our calculations with $\langle n\rangle \le 100$ this manifests as a limited fidelity to an ideal single-mode cat and a reduced purity, even though a clear two-lobe structure and quantum interference fringes are visible in phase space. For the case $\langle n\rangle=100$ we find a purity (trace of $\rho^2$) $\simeq 0.15$ and a fidelity $F\simeq 0.26$ (with respect to the closest ideal cat), while the state remains nonclassical as witnessed by the Wigner negativity. 
    This is also indicated by the probability distributions of quadratures: $p(X)=\int W(X,Y)dY$ with manifestation of quantum interference and $p(Y)=\int W(X,Y)dX$ with two probability zones.
	
	The non-monotonic behavior of the mean pump photon number is due to  the reversible nature of the interaction between the pump mode and second harmonic. The values of the interaction time optimal for the formation of states (the moment of the maximum of mean photon number, Fig.~\ref{fig:fig2}) were calculated in the quantum evolution only for the initial number of photons $\leq$ 100.
    To estimate this time in a case that is closer to real conditions, we used a classical simulation method, where the quantum behavior was simulated by the trajectories of classical waves fluctuating at input with proper values.
    	
	\section{Classical Simulation of Quantum Solutions}
	
	The following “classical simulation” can be interpreted as a truncated-Wigner-type approach: quantum fluctuations of the input coherent (pump) and vacuum (second-harmonic) states are represented by sampling the corresponding Wigner distribution, followed by deterministic propagation of each trajectory. Classical field trajectories in the second harmonic generation process were calculated using equations with the classical Hamiltonian function of type (1)
	\begin{align}
	\hat{H} / i\hbar \iff &H_{\text{clas}} = ig \left( a_1^{*2} a_2 + a_2^* a_1^2 \right),\\
	&\frac{da_1}{dt} = 2ig a_2 a_1^*,\\
	&\frac{da_2}{dt} = ig a_1^2.
	\end{align}
		These equations correspond to the equations for the field operators in the Heisenberg interaction picture. Random amplitudes simulating quantum fluctuations of coherent and vacuum states of the input waves were taken as initial conditions:
	\begin{eqnarray}
	&\alpha_1 = \alpha + x_1 + iy_1, \text{  } \alpha_2 = x_2 + iy_2,\\
    	\langle x_{1,2} \rangle &= \langle y_{1,2} \rangle = 0, \text{  }\langle \Delta x_{1,2}^2 \rangle = \langle \Delta y_{1,2}^2 \rangle = 1/4,
	\end{eqnarray}
		where the distributions of random values \(x_{1,2}\) and \(y_{1,2}\) are Gaussian. A rigorous justification of the correspondence between classical trajectories and quantum behavior of wave functions is beyond the scope of this article and will be published separately. Here we demonstrate nearly perfect coincidence of the results of quantum and classical calculations for the dynamics of the mean photon number of the pump wave (Fig.~\ref{fig:fig1}). For large photon numbers $\langle n \rangle \geq 100$ we performed classical simulation in order to find the time $gt_{\text{max}}$, where the mean number of photons in pump mode reaches maximum value after first minimum (Fig.~\ref{fig:fig3}). We use this time for simulation of pump field distributions (see below). The calculated values are approximated by formula
	\begin{equation}
	gt_{\text{max}} \approx \frac{2.8\pm{0.1}}{\langle n \rangle^{0.44\pm{0.005}}}.\label{eq:gtmax}
	\end{equation}
	  \begin{figure}[htbp]
		\centering
\includegraphics[width=0.5\textwidth]{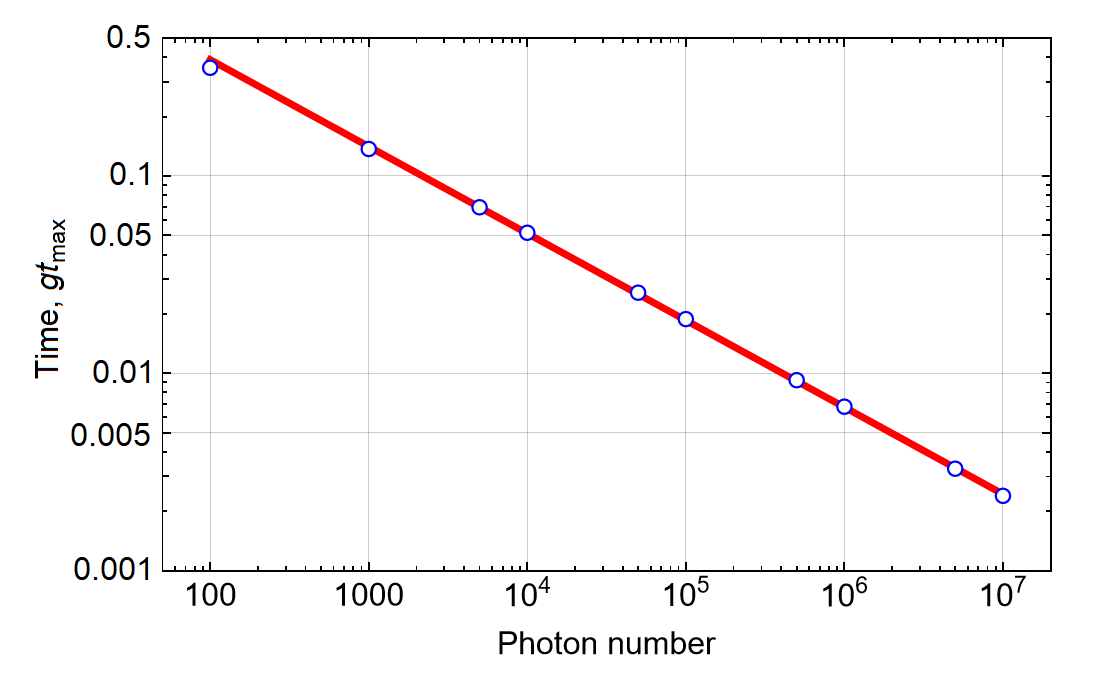}
		\caption{Time of first maximum \(gt_{\text{max}}\) as a function of the initial mean pump wave photon number (circles) and approximating line according to formula (\ref{eq:gtmax}).}
		\label{fig:fig3}
	\end{figure}
Unexpectedly, the mean pump wave photon number at the moment of first maximum $gt_{\text{max}}$ for all calculated cases turned out to be at the level of \((64 \pm 1)\%\) of the initial photon number. Notably, a similar conversion efficiency (approximately 63.2\%) was found optimal for the formation of cat-like states during depleted-pump parametric down-conversion \cite{Singh2024arXiv,Singh2026}.

\begin{figure*}[htbp]
		\centering
		\includegraphics[width=\linewidth]{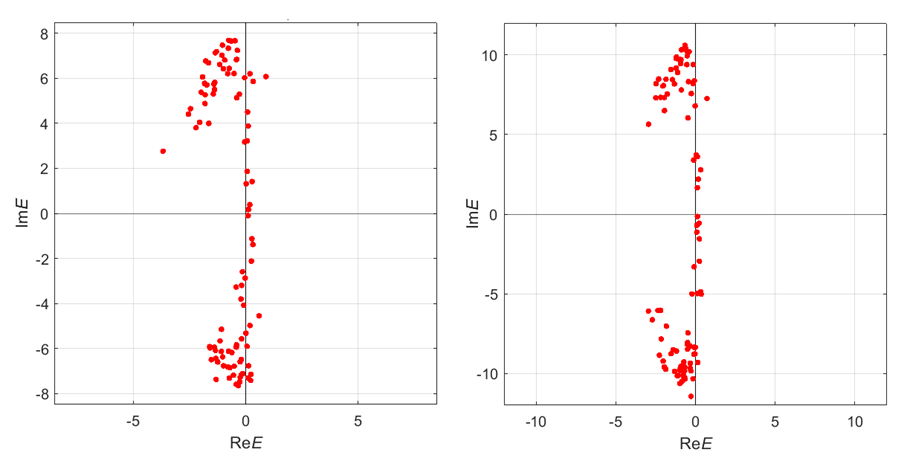}
		\caption{Distribution of resulting pump field amplitudes in the classical picture. Points correspond to individual realizations of solutions with random initial values; {\it{a}} — initial mean photon number $\langle n \rangle = 50$, time \(gt_{\text{max}} = 0.463\), {\it{b}} — initial mean photon number $\langle n \rangle = 100$, time \(gt_{\text{max}} = 0.352\) (see Fig.\ref{fig:fig1}).}
		\label{fig:fig4}
	\end{figure*}

The classical simulation method was applied for calculating the distributions of complex field amplitudes of the pump wave at the moment of maximum \(gt_{\text{max}}\) (Fig.~\ref{fig:fig4}). Each point in Fig.~4 corresponds to the final pump wave amplitude; 100 samples are presented. Data demonstrate excellent quantitative agreement with main features of Wigner functions Fig.~2, except zones with negative quasiprobabilities.

\section{Conclusions}
We have shown that the target quantum state close to SC state can be achieved in the process of second harmonic generation when the pump wave reaches the maximum energy after depletion. Such a regime of deep depletion and revival of pump energy had not been realized in experiments with $\chi^{(2)}-$ nonlinearity. However, as the technique for generating the optical second harmonic improves, such a regime may become feasible. High efficiency second harmonic generation has been has been experimentally achieved with straight waveguides based on lithium niobate, where the conversion coefficients reach $(0.01$--$0.1)\,\mathrm{mW}^{-1}$ at pump wavelengths $\lambda \approx 1.5\,\mu\mathrm{m}$ with waveguide lengths about few mm \cite{Wang2018, Zhao2020, Boes2021, He2024, Yang2024}. In such devices, the effect of pump depletion has been observed \cite{Chen2024}. In the range of $\mu\mathrm{W}$ pump powers, efficient harmonic generation has been achieved in lithium niobate microring resonators \cite{McKenna2022, Cheng2023}, also with effects of pump depletion \cite{Lu2019}.
As the calculations show, the method of classical simulation turns out to be consistent in estimating the quantum behavior of light in nonlinear optical processes. With this method, it is possible to estimate the quantum evolution of light waves with large numbers of photons $(\langle n \rangle = 100)$, when quantum calculations are practically impossible due to the limited power of modern personal computers. Although there is no rigorous justification for the validity limits of classical simulation, the method is attractive due to its clarity and ease of implementation (see, for example, \cite{Leuchs2021}).
In addition, classical simulation allows searching for interaction regimes with unusual quantum properties of light waves. In the case of second-harmonic generation that we have considered, the splitting of the {\it{Y}}-quadrature probability density into two zones (Fig.~\ref{fig:fig2}a) clearly indicates the time of evolution when quantum interference can occur. We expect that classical simulation will be a reliable method for describing quantum Gaussian states where the Wigner quasi-probability is positive.

\section{ACKNOWLEDGMENTS}
R.S. is grateful to A.E. Teretenkov for insightful comments.

\appendix
\section*{Appendix 1}

This section provides an estimate of the conditions for generating light in the Schrödinger cat (SC) state when radiation passes through a Kerr medium. The interaction Hamiltonian and the evolution operator in this case have the form \cite{Kitagawa1986}:
\begin{align}
\hat{H} &= \frac{n_2 (\hbar \omega)^2}{2 n_0 \tau_{\mathrm{coh}} \sigma}\hat{n}^2,\\
\hat{U} &= \exp(-i\hat{H}t/\hbar)
        = \exp\!\left(\frac{i\Phi_{\mathrm{NL}}}{2\langle \hat{n} \rangle}\hat{n}^2\right).
\end{align}
where $n_2$ is the nonlinearity coefficient of the medium; $\Phi_{\text{NL}} = \omega n_2 I z/c$ is the noninear phase, which plays the role of evolution time; $\tau_{\text{coh}}$ is the coherence time of the radiation; $\sigma$ is the effective cross-section of the light beam; $n_0$ is the refractive index of the medium. As shown in \cite{Miranowicz1990}, due to the accumulation of nonlinear phase in the medium, the light undergoes a change in its quantum state, sequentially transitioning from the initial coherent state to superpositions of many coherent states: ..., four, three, and finally two coherent states, i.e., the SC state. The required phase accumulation in such a process occurs only with sufficiently high radiation intensity. 
Indeed, for \( \Phi_{\text{NL}} = \pi \langle n \rangle \) we have:
\begin{align}
\vert \psi\rangle &= \hat{U}\vert \alpha\rangle
= e^{-\vert\alpha\vert^{2}/2} \sum_{n=0}^{\infty} e^{i\pi n^{2}/2} \frac{\alpha^{n}}{\sqrt{n!}} \vert n\rangle \nonumber\\
&= e^{-\vert\alpha\vert^{2}/2} \Big( \sum_{m=0}^{\infty} \frac{\alpha^{2m}\vert 2m\rangle}{\sqrt{(2m)!}}
 + i \sum_{m=0}^{\infty} \frac{\alpha^{2m+1}\vert 2m+1\rangle}{\sqrt{(2m+1)!}}  \Big) \nonumber\\ \nonumber
&= \frac{1}{2} \bigl( \vert \alpha\rangle + \vert -\alpha\rangle\bigr) + \frac{i}{2} \bigl( \vert \alpha\rangle - \vert -\alpha\rangle\bigr) \\
&= \frac{e^{i\pi/4}}{\sqrt{2}} \Bigl( \vert \alpha\rangle + e^{-i\pi/2} \vert -\alpha\rangle \Bigr).
\end{align}
This state is neither even SC nor odd SC (neither cat nor kitten). The photon number distribution here is Poissonian, in contrast to the photon number distributions of the even and odd SC states.

\appendix
\section*{Appendix 2}

The instability of the pump phase during second harmonic generation can be demonstrated using classical equations. During the interaction process, the slow amplitudes of the pump wave \( A_p \) and the second harmonic \( A \) change as follows:
\begin{align}
\frac{dA_P}{dt} &= 2ig A A_P^*, \\
\frac{dA}{dt} &= ig A_P^2,
\end{align}
where \( g \) is the interaction parameter (\( g \propto \chi^{(2)} \)) and perfect phase matching is implied. The bifurcation of the pump wave phase occurs at the initial stage of interaction in the presence of a nonzero seed of the second harmonic wave \( A(0) \neq 0 \); such a seed simulates the quantum vacuum state of the wave at the entrance to the medium. For simplicity, we assume the pump wave amplitude \( A_p(0) \) to be real. Let us separate the seed into the part \( a' \) that is in phase with the pump and the orthogonal part \( a'' \): \( A(0) = a' + ia'' \). Then, for a small evolution time $\Delta t$ the pump amplitude becomes:
\begin{eqnarray}
A_P(\Delta t) \approx A_P(0) + 2ig A_P^*(0)a'\Delta t - 2gA_P^*(0) a'' \Delta t.    
\end{eqnarray}
The inphase part of seed $a'$ alters the pump wave phase either clockwise or counter-clockwise depending on the sign of $a'$. Since the vacuum state possesses components of both signs both phase evolutions of pump field occur, i.e., the pump wave experiences phase instability. In the quantum description, this instability manifests itself in the growth of the phase spread, which at a certain moment leads to the separation of the phase into two regions resembling the SC state. Note, that similar speculation is not valid for the influence of orthogonal part of seed $a''$, because the growth of the pump amplitude due to the vacuum seed is not allowed.

\bibliographystyle{apsrev4-2}
\bibliography{references} 
\end{document}